\documentclass{JHEP3}

\usepackage{graphicx}
\usepackage{multicol}
\usepackage{bbm}
\usepackage{amsmath}
\usepackage{amssymb}
\usepackage{euscript}
\usepackage{array}
\usepackage{amsfonts}
\usepackage{mathrsfs}

\newcommand{\ud}{\mathrm{d}} 

\title{
A new approach to cosmological perturbations in \boldmath $f(R)$ models
}

\author{
Daniele Bertacca$^{a,b,c}$, Nicola Bartolo$^{a,b}$, Sabino Matarrese$^{a,b}$\\
$^a$ Dipartimento di Fisica Galileo Galilei, Universit\`{a} di Padova, via F. Marzolo, 8 I-35131 Padova, Italy\\
$^b$ INFN Sezione di Padova, via F. Marzolo, 8 I-35131 Padova, Italy\\
$^c$ Institute of Cosmology \& Gravitation, University of Portsmouth, Dennis Sciama Building, Portsmouth, PO1 3FX, United Kingdom\\
E-mails: \email{daniele.bertacca@pd.infn.it}, \email{nicola.bartolo@pd.infn.it}, \email{sabino.matarrese@pd.infn.it}
}

\abstract{
We propose an analytic procedure that allows to determine quantitatively the 
deviation in the behavior of cosmological perturbations between a given $f(R)$ modified gravity model and 
a $\Lambda$CDM reference model. 
Our method allows to study structure formation in these models from the largest 
scales, of the order of the Hubble horizon, down to scales deeply inside the Hubble radius, without 
employing the so-called ``quasi-static" approximation.   
Although we restrict our analysis here to linear perturbations, our technique is completely general and can 
be extended to any perturbative order.
} 
\keywords{alternative theory of gravity;  cosmological perturbations; analytical method}

\begin{document} 

\section{Introduction}

During the last decade,  independent observational data such as type-Ia Supernovae (SNIa) \cite{Perlmutter:1998np, Riess:1998cb, Riess:1998dv, Amanullah:2010vv} and Cosmic Microwave Background (CMB) 
\cite{Larson:2010gs, Komatsu:2010fb} and Baryonic Acoustic Oscillations \cite{Eisenstein:2005su, Percival:2007yw, Percival:2009xn} suggest that two dark
components govern the dynamics of the Universe. They are the Dark Matter (DM), thought to be
the main responsible for structure formation, and a non-zero cosmological constant $\Lambda$ 
(see, e.g. ref.~\cite{Weinberg:1988cp}) or a dynamical dark energy (DE) component, that is supposed to drive the observed 
cosmic acceleration \cite{Tsujikawa:2010sc, Copeland:2006wr}.

The standard cosmological model $\Lambda$CDM provides a good fit to observations, but it describes the Universe by means of two unknown components, which represent 96\% of the total energy density. 
However, it should be recognised that, while some form of Cold Dark Matter (CDM) is independently expected to exist within any modification of the Standard Model of high energy physics, 
the really compelling reason to postulate DE has been the discovery that the Universe is experiencing a phase of accelerated expansion.
This could be theoretically unsatisfactory and, recently, alternative models have been proposed with respect to the class of DE models.  
In particular, many of them are based on modifications of gravity at large distances. 
For example, scalar-tensor theories \cite{Brans:1961sx, Amendola:1999qq, Uzan:1999ch, Chiba:1999wt, Bartolo:1999sq, Perrotta:1999am, Boisseau:2000pr}, 
Brane-World models (see e.g the review \cite{Maartens:2010ar} and refs. therein),  Galileon models \cite{Deffayet:2009wt, Deffayet:2009mn, DeFelice:2010pv, 
DeFelice:2010nf, Chow:2009fm, Silva:2009km, Kobayashi:2009wr}, Gauss-Bonnet gravity \cite{Carroll:2004de, Mena:2005ta, Nojiri:2005vv} and other scenarios (see, 
for example, the review \cite{Tsujikawa:2010sc} and refs. therein).

An interesting class of modified gravity models is represented by  
$f(R)$ theories (see  the reviews \cite{Faraoni:2004pi, Nojiri:2006ri, Sotiriou:2008rp, DeFelice:2010aj, Tsujikawa:2010zza, Nojiri:2010wj, Capozziello:2011et} and refs. therein\footnote{See also 
\cite{Raccanelli:2011pu, Hojjati:2011ix, Zhao:2011cu, Dossett:2011zp, Paschalidis:2011ww, Linder:2011hi, Motohashi:2011wy, Zhao:2010qy, Ferraro:2010gh, Motohashi:2010sj, Dunsby:2010wg, Baghram:2010mc, Schmidt:2010jr,  Zhao:2010dz, Motohashi:2010tb, Capozziello:2009nq, Schmidt:2009am, Carloni:2009gp, Motohashi:2009qn, Ananda:2008tx, Carloni:2007yv, Carloni:2004kp} for other recent contributions to this area.}), whose Lagrangian density is simply defined by an arbitrary function of the Ricci scalar $R$. 
These Lagrangians were proposed for the first time in connection with Inflation in the early Universe \cite{Starobinsky:1980te} and, only recently, they have been 
used in the context of DE models, to explain the present-day cosmic acceleration
(see e.g. \cite{Capozziello:2003tk, Carroll:2003wy, Nojiri:2003ft}). 

In general, there are two ways to study these models \cite{Bertschinger:2008zb}: 1) specifying directly the type of Lagrangian that satisfies cosmological and local gravity constraints 
(e.g.  \cite{Starobinsky:2007hu, Hu:2007nk} and see also \cite{DeFelice:2010aj})
2) through a parametrized post-Newtonian framework or similar (see e.g. Refs.\ \cite{Bertschinger:2006aw,  Hu:2007pj, Caldwell:2007cw, Amendola:2007rr, Amin:2007wi, 
Linder:2007hg, Bertschinger:2008zb, Hu:2008zd, Ferreira:2010sz}), 
to  provide a scale-dependent parameterization of cosmological perturbations (see also \cite{BuenoSanchez:2010wd}). 

Recently, it was found that, in viable $f(R)$ models (see e.g. \cite{Song:2006ej, Bean:2006up, Starobinsky:2007hu, 
Hu:2007nk, Pogosian:2007sw, Zhang:2007ne, Sawicki:2007tf,  Nojiri:2007as, Cognola:2007zu, Motohashi:2009qn, DeFelice:2010aj}) when the superhorizon long-wavelength limit is taken 
(i.e. when spatial gradients may be neglected in the equations), the evolution of metric and density perturbations of a Robertson-Walker 
background can be described by the Friedmann equation and energy-momentum conservation 
\cite{Bertschinger:2006aw, Hu:2007pj, Starobinsky:2007hu, Bertschinger:2008zb, Hu:2008zd}.  In addition,
the  viable $f(R)$ models usually need to be close to the $\Lambda$CDM model during the matter dominated
epoch in order to satisfy cosmological constraints, local gravity constraints (see \cite{Hu:2007nk, DeFelice:2010aj, PH,  Lombriser:2010mp, Seljak}) and galactic constraints \cite{Hu:2007nk}.  Motivated by these results, in this paper, we propose an analytical method which allows to describe  both the background evolution and cosmological perturbations in 
$f(R)$ modified gravity models. This approach is completely different from previous ones. 
Indeed, through this analytic technique we can determine quantitatively the deviation in the behavior of cosmological perturbations between a given 
$f(R)$ model and a $\Lambda$CDM reference model.
Moreover, our treatment is general in that all the results depend only on the initial conditions that characterize the type of $f(R)$ model we are studying. 
Finally, our approach allows to study structure formation in these models from the largest 
scales, of the order of the Hubble horizon, down to scales deeply inside the Hubble radius, without 
employing the so-called ``quasi-static" approximation (see also \cite{delaCruzDombriz:2008cp, delaCruzDombriz:2009xk}).   
Although we restrict our analysis here to linear perturbations, the technique can be applied to linear and weakly non-linear scales  and can 
be extended to any perturbative order (see also \cite{Tatekawa:2008bw, Bernardeau:2011sf}).
Of course, in order to study these models at non linear scales another approach must be adopted (for example, see \cite{DeFelice:2010aj}). 

The rest of the paper is organised as follows. In Section~\ref{Sec:PreRel} we present the basic equations describing the background and the 
perturbative evolution of a generic $f(R)$ gravity model. In Section~\ref{Technique} we describe our approach, analyzing both the background 
evolution and the first-order perturbation equations in the synchronous gauge and, in Section~\ref{poissongauge}, we introduce  the evolution of cosmological perturbations in the Poisson gauge. In Section \ref{quasi-static} we recover the  ``quasi-static" approximation  and Section \ref{MPS} we compute the matter power spectrum with our approach.
Section \ref{conclusions} is devoted to our conclusions. Appendix \ref{A} and Appendix \ref{B} are devoted to the evolution of linear perturbations in the reference $\Lambda$CDM model, in the synchronous and Poisson gauge, respectively.\\
Throughout the paper we use $G = c = 1$ units and the $(-,+,+,+)$ signature for the metric. Greek indices run over $\{0,1,2,3\}$, denoting space-time coordinates, whereas Latin indices run over $\{1,2,3\}$, labelling spatial coordinates.

\section{Preliminary relations}
\label{Sec:PreRel}

Let us consider the following action in $f(R)$ gravity  (see e.\ g.\  \cite{Starobinsky:1980te, Capozziello:2003tk, Carroll:2003wy, Nojiri:2003ft}; see also the reviews 
\cite{Faraoni:2004pi, Sotiriou:2008rp, DeFelice:2010aj,Tsujikawa:2010zza} and refs. therein):
\begin{equation}\label{action}
S=S^{\rm (GR)}+S^{\rm (m)}=\frac{1}{16\pi}\int d^4 x\sqrt{-g}\,f(R) +  \int d^4 x\sqrt{-g}\, {\cal L}_{\rm m}[\Psi_{\rm m},g_{\mu\nu}]  \;.
\end{equation}
In this case $f(R)$ is a general function of the Ricci scalar, $R$,
and ${\cal L}_{\rm m}$  is the matter Lagrangian and $\Psi_{\rm m}$ are the matter fields. 
Defining $\phi \equiv \partial f/\partial R$,  $S^{\rm (GR)}$ can be cast in the form of Brans-Dicke (DB) theory \cite{Brans:1961sx} with a potential for the scalar field $\phi$ 
\cite{Faraoni:2004pi, Sotiriou:2008rp, DeFelice:2010aj,Tsujikawa:2010zza}. 

In particular,
\begin{equation}\label{action}
S^{\rm (GR)}=\frac{1}{16\pi} \int d^4 x\sqrt{-g}\, \left[\phi R- V(\phi) \right]  \;,
\end{equation}
where $V(\phi)=R\phi-f(R)$.
The field equations obtained from varying the action with respect to $g_{\mu\nu}$ are
\begin{equation} \label{fieldeq}
\phi G_{\mu \nu} - 8\pi T_{\mu \nu}^{\, \phi}=8 \pi T_{\mu \nu}^{\rm \, m} \;,
\end{equation}
where $G_{\mu \nu}=R_{\mu \nu} -(1/2) R g_{\mu\nu}$, 
\begin{equation}
 8\pi T_{\mu \nu}^{\, \phi}= \nabla_\mu \nabla_\nu \phi - \Box \phi g_{\mu \nu}- \frac{1}{2} V(\phi) g_{\mu \nu}\, ,
 \end{equation}
and $T_{\mu\nu}^{\rm \, m} $ is the energy-momentum tensor  of a perfect fluid, i.e.
\begin{equation}
T_{\mu\nu}^{\rm \, m} = - \frac{2}{\sqrt{-g}}\frac{\delta S^{\rm (m)}}{\delta g^{\mu \nu }}=(\rho_{\rm m}+ p_{\rm m})u_{\mu} u_{\nu} + p_{\rm m} g_{\mu\nu}\;.
\label{perfectfluid}
\end{equation} 
The vector $u^{\mu}$ is the fluid rest-frame four-velocity, $\rho$ is the energy density and $p$ the isotropic pressure.
In this case the equation of motion of the scalar field is simply
\begin{equation}
\label{eqmotion}
R=\frac{\ud V}{\ud \phi}\;.
\end{equation}
Finally by taking the trace of  Eq.\ (\ref{fieldeq}) and using Eq.\ (\ref{eqmotion}), we obtain the dynamics of the scalar field for a given matter source
 \begin{equation}\label{trace-fieldeq}
 \Box \phi + \frac{1}{3}\left(2V-\phi \frac{\ud V}{\ud \phi} \right) = \frac{8 \pi}{3} T^{\rm \, m} \;,
\end{equation}
where $T^{\rm \, m}=g^{\mu\nu} T_{\mu\nu}^{\rm \, m}$.

When considering the background cosmological evolution, we take the metric to be of the flat Friedmann-Lema$\rm \hat{\i}$tre-Robertson-Walker (FLRW) 
form, $ds^2 = a^{2}(\eta)(-d\eta^{2} + d{\bf x}^2)$, where $\eta$ is the comoving time and $a(\eta)$ the scale factor. Then the gravitational field equations, the scalar 
field equation of motion and the continuity equation become
\begin{eqnarray}\label{back-00}
A \mathcal{H}^2 &=& \frac{8 \pi a^2}{3} \rho_{\rm m}^{(0)}+(A-\varphi) \mathcal{H}^2 + (A-\varphi)'  \mathcal{H} +\frac{a^2}{6}V(\varphi) \;,\\ \label{back-ij}
A \left( 2\mathcal{H}' + \mathcal{H}^2 \right)&=&- 8 \pi a^2 p_{\rm m}^{(0)} + (A-\varphi) \left( 2\mathcal{H}' + \mathcal{H}^2 \right) + (A-\varphi)''+ 
(A-\varphi)'  \mathcal{H} + \frac{a^2}{2} V(\varphi) \;, \nonumber \\ \\
\label{back-eqmotion}
\mathcal{H}' + \mathcal{H}^2&=&\frac{a^2}{6} \frac{\ud V}{\ud \varphi}  \;, \\ \label{conteq}
 \rho_{\rm m}^{(0)}\,'  &+& 3\mathcal{H} (p_{\rm m}^{(0)} + \rho_{\rm m}^{(0)})=0\, ,
\end{eqnarray} 
where $A$ is a suitable constant (for its physical meaning see Ref.\ \cite{DeFelice:2010aj}), $\mathcal{H}=a'/a$ and primes indicate differentiation w.r.t. $\eta$. 
Note that $\phi^{(0)}=\varphi$, $p_{\rm m}^{(0)}$ and $\rho_{\rm m}^{(0)}$ are respectively the background scalar field, matter pressure and matter energy density.
From Section~\ref{FLRW} onwards we will set $A=1$.

Now taking into account only the non-relativistic matter, i.e. $p_{\rm m}=0$,  let us consider the metric of a flat FLRW Universe with small perturbations. 
In particular we want to study the evolution of the scalar perturbations  in  the synchronous gauge (see also  Ref.\ \cite{Bean:2006up}).  In this case the line-element is written in the 
form $ds^2=a^2 \{- d\eta^2 +[ (1-2\psi^{(1)})\delta_{ij}+D_{ij}\chi^{(1)} ) ] dx^i dx^j \}$, where $D_{ij}=\partial_i \partial_j - (1/3) \delta_{ij} \nabla^2$ is a 
trace-free operator. Let us allow for small inhomogeneities of the scalar field, $\phi(\eta,\mbox{\bf x})= \varphi(\eta)+\varphi^{(1)}(\eta,\mbox{\bf x})$.
Perturbing  Eq.\ (\ref{fieldeq}), we get
\begin{eqnarray}
\label{pert-00}
&& 3\left(2\mathcal{H}+\frac{\varphi'}{\varphi}\right)\psi^{(1)}\,'  -\left(2 \nabla^2\psi^{(1)} +\frac{1}{3} 
\nabla^2\nabla^2 \chi^{(1)} \right) = \frac{8 \pi a^2}{\varphi} \rho_{\rm m}^{(0)} \left(\frac{\varphi^{(1)}}{\varphi}-\delta^{(1)}\right)  \nonumber \\
&& + 3 \mathcal{H} \frac{\varphi^{(1)}\,'}{\varphi} -3\mathcal{H} \frac{\varphi'}{\varphi} \frac{\varphi^{(1)}}{\varphi} -  
\frac{\nabla\varphi^{(1)}}{\varphi} + \frac{a^2}{2} \left(\frac{V}{\varphi}-\frac{\ud V}{\ud \varphi}\right)\frac{\varphi^{(1)}}{\varphi} \;,  \\  \nonumber \\
\label{pert-0i}
&&2\psi^{(1)}\,' +\frac{1}{3} \nabla^2 \chi^{(1)}\,' = \frac{\varphi^{(1)}\,'}{\varphi} - \mathcal{H} \frac{\varphi^{(1)}}{\varphi} \;, \\  \nonumber \\
\label{pert-ij}
&&\chi^{(1)}\,''  + \left(2\mathcal{H} + \frac{\varphi'}{\varphi}\right)  \chi^{(1)}\,' + 2 \psi^{(1)} +\frac{1}{3} \nabla^2 \chi^{(1)} = 2 \frac{\varphi^{(1)}}{\varphi} \;,
\end{eqnarray} 
where $\delta^{(1)}=(\rho_{\rm m}^{(1)}-\rho_{\rm m}^{(0)})/\rho_{\rm m}^{(0)}$. From  Eq.\ (\ref{trace-fieldeq}), we obtain 
\begin{eqnarray}
 \label{pert-trace}
\varphi^{(1)}\,'' +2 \mathcal{H} \varphi^{(1)}\,'  - \nabla^2 \varphi^{(1)}  - 3 \varphi' \psi^{(1)}\,' = \frac{8 \pi a^2}{3}   \rho_{\rm m}^{(0)} \delta^{(1)} -  
\frac{a^2}{3} \left(\varphi \frac{\ud^2 V}{\ud \varphi^2} - \frac{\ud V}{\ud \varphi}\right)  \varphi^{(1)} \;. \nonumber \\ 
\end{eqnarray} 
Finally, at the linear order,  Eqs.\ (\ref{eqmotion}) and (\ref{conteq}) become
\begin{eqnarray}
\label{pert-eqmotion}
&&-6 \psi^{(1)}\,'' -18 \mathcal{H} \psi^{(1)}\,' + 4 \nabla^2 \psi^{(1)} + \frac{2}{3}  \nabla^2\nabla^2 \chi^{(1)} = a^2  \frac{\ud^2 V}{\ud \varphi^2}  
\varphi^{(1)} \;,\\  \nonumber \\
\label{pert-conteq}
&&\delta^{(1)}\,'=3\psi^{(1)}\,' \;.
\end{eqnarray} 

\section{Description of the analytical approach}
\label{Technique}

Motivated by the fact  that, in viable $f(R)$ models \cite{DeFelice:2010aj}, 
1) the cosmic scale-factor $a(\eta)$ can be described by the Friedmann equation 
with background expansion of the Universe close to the $\Lambda$CDM model and 2)  gravity can be described by a classical 
four-dimensional metric theory having a well-defined infrared limit \cite{Song:2006ej,Bertschinger:2006aw, Bean:2006up, 
Starobinsky:2007hu, Pogosian:2007sw, Bertschinger:2008zb} (i.e. in the modified gravity models, the long-wavelength perturbations 
obey the same constraints as they do in general relativity \cite{Bertschinger:2006aw, Hu:2007pj, Starobinsky:2007hu, Bertschinger:2008zb}),
let us make the following ans$\ddot{\rm a}$tz: 

{\it i)} at the background level, the scalar field can be described in the following way
\begin{equation}
\varphi=  \bar{\varphi} + \epsilon A \xi +... \;,
\end{equation}
where we introduced a suitable perturbative parameter $|\epsilon| \ll 1$;

{\it ii)} at the first order level 
\begin{eqnarray}
\psi^{(1)} &=& \psi^{(1,0)} +\epsilon A \, \psi^{(1,1)} + ... \;, \\
\chi^{(1)} &=& \chi^{(1,0)} +\epsilon A \, \chi^{(1,1)} + ... \;, \\
\delta^{(1)} &=& \delta^{(1,0)} +\epsilon A \, \delta^{(1,1)} + ... \;,
\end{eqnarray}
and 
\begin{equation}
\varphi ^{(1)} = \xi^{(1,0)} +\epsilon A \xi^{(1,1)}+\frac{\epsilon^2}{2} A\xi^{(1,2)}+... \;,
\end{equation}
where in the double superscript $(i,j)$ the left index $i$ refers to the order in standard perturbation theory (e.g. $i=1$ 
denotes linear theory), while the right index $j$ refers to the order $\epsilon^j$ of the deviation of our modified gravity 
model w.r.t. $\Lambda$CDM. 
In particular,  setting\footnote{ If we consider $\bar{\varphi}$ non-constant then the background could change 
completely. Indeed, as we will see from Eqs.\  (\ref{back-00_standard}) and (\ref{beck-ij_standard}), when $\bar{\varphi} =1$ and $A=1$, the value of $\mathcal{H}$ does not depend on $\varphi$ and, obviously, on $\epsilon \xi$.Therefore the background is the same of the $\Lambda$CDM model for $\epsilon \xi \ll 1$.} $\bar{\varphi} =1$, $A=1$ and 
$\xi^{(1,0)}=0$ we recover the $\Lambda$CDM model when  $\epsilon \to 0$. 
In other words,  all the equations previously obtained will be also iteratively perturbed by the parameter $\epsilon$. 
This approach allows us to understand the relevance of the various terms that come from modified gravity 
in the description of the large scale structure of the Universe. Let us  stress that this prescription is completely general
and can be used to describe any viable $f(R)$ theory on cosmological scales.

In the next two subsections we will analyze in detail with this technique the background 
and the cosmological perturbations. Specifically we will expand  all the equations in section \ref{Sec:PreRel} up to 
order $\epsilon$.

\subsection{FLRW background equations}
\label{FLRW}

In this case, by construction,  $V(\bar{\varphi})=V_0$ is a constant (specifically, it is of the order of the cosmological constant), $\bar{\varphi} =1$, $A=1$ and 
$\rho_{\rm m}^{(0)}$ is unperturbed with respect to $\epsilon$. In this case from Eq.\ (\ref{back-00}) we get
\begin{eqnarray}
\label{back-00_standard}
\mathcal{H}^2=\frac{8 \pi a^2}{3} \rho_{\rm m}^{(0)}+ \frac{a^2}{6}V_0 \;, \\
\label{back-00_xi}
\mathcal{H} \xi ' + \mathcal{H}^2  \xi =  \frac{a^2}{6}V_1\;,
\end{eqnarray}
where we have defined the expansion of $V(\varphi)$ with respect to $\epsilon$ as $V(\varphi)=V_0 + \epsilon V_1(\xi(\eta))$. Instead, from Eq.\ (\ref{back-ij}) we obtain
\begin{eqnarray}
\label{beck-ij_standard}
 \left( 2\mathcal{H}' + \mathcal{H}^2 \right) =  \frac{a^2}{2} V_0\;, \\
 \label{back-ij_xi}
 \xi '' + \mathcal{H} \xi' + \left( 2\mathcal{H}' + \mathcal{H}^2 \right)\xi= \frac{a^2}{2} V_1 \;.
 \end{eqnarray}
We note immediately that we retrieve the field equations of general relativity at the background level. Now from Eqs.\ (\ref{back-00_xi}) and (\ref{back-ij_xi}) we find the equation of motion for $\xi$
\begin{equation} 
\label{Eqxi}
 \xi ''  -2 \mathcal{H}  \xi' + 2\left(\mathcal{H}' - \mathcal{H}^2  \right) \xi =0 \;.
 \end{equation}
Notice that this field equation describes the dynamics of $\xi$ as a free field, without any source term proportional to the matter fields. 

At this point, knowing that the Cauchy problem is well-formulated for metric $f(R)$ gravity in the presence of matter \cite{LanahanTremblay:2007sg}, 
the initial conditions for $\xi$ and $\xi'$ are crucial because they characterize the type of $f(R)$ model that we are studying
and the validity of the approach that we are using\footnote{This translates to imposing the conditions on  $\partial f/ \partial R$ and Ê$\partial^2f/\partial R^2$ (or $B$ defined by \cite{Song:2006ej}) at a specific time (for example, at recombination or today). In particular, let us look at Eq.\ (3) of Ref.\ \cite{Ferraro:2010gh} (see also  \cite{Hu:2007nk}). In this paper the authors need simply to set two parameters, namely $f_{R_0}$ and $n$ in order to study one class of viable $f(R)$ models (see Section \ref{MPS}).}. 
Specifically, assuming that for $\eta<\eta_{\rm rec}$, where $\eta_{\rm rec}$ is some epoch when the Universe is matter dominated and radiation is negligible (usually at the recombination epoch),
$f(R)$ gravity should be the same as general relativity. Thus we can choose at recombination $\xi(\eta_{\rm rec})=0$ and $| \xi(\eta_{\rm rec})' | \ll 1$.  
In other words, we have to set both  $\epsilon$  and $| \xi(\eta_{\rm rec})' |$ ``small enough" to allow only at late times a different  solution of the perturbation evolution w.r.t. a $\Lambda$CDM
model.

Let us make another comment. For the stability at high curvature of perturbations of the scalar field the sign of $\epsilon$ and $ \xi'$ is crucial. Indeed, from Eq.\ (\ref{pert-trace}) 
we must assume that  $(a^2/3) \left[\varphi (\ud^2 V/\ud \varphi^2) - (\ud V/\ud \varphi)\right]>0$  \cite{Song:2006ej, Starobinsky:2007hu, Pogosian:2007sw, DeFelice:2010aj}. Now, taking into account that 
\begin{equation}
 \frac{\ud^2 V}{\ud \varphi^2} = \frac{6/a^2 }{ \epsilon \xi '} \left(\mathcal{H}'' -2\mathcal{H}^3 \right) = 8 \pi \frac{\rho_{\rm m}^{(0)}\,'}{\epsilon \xi '}=- 24 \pi \frac{\rho_{\rm m}^{(0)}}{\epsilon (a \ud \xi/ \ud a)} \;,
\end{equation}
for $\epsilon \to 0$, we have to impose that $\epsilon (a \ud \xi/ \ud a) <0$. Moreover, analyzing Eq. (\ref{back-ij_xi})  we note that if $\xi'<0$($>0$) 
then $\xi <0$ ($>0$). Therefore, assuming for simplicity $\epsilon>0$  we have to set $ (a \ud \xi/ \ud a) <0$.

Now, in order to better understand the behaviour of  $\xi$, let us rewrite Eq.\  (\ref{Eqxi}) in a slightly different way. Indeed, using Eqs.\ (\ref{back-00_standard}) 
and (\ref{beck-ij_standard}), defining $V_0 = 16 \pi \rho_\Lambda$ and $\nu=\rho^{(0)}_{\rm 0 \, m}/\rho_\Lambda=(\Omega_{0 \,  \rm m}/\Omega_{0 \,  \Lambda})$, where $\rho_{\rm m}^{(0)}= \rho^{(0)}_{\rm 0 \,  m} a^{-3}$ and\footnote{$a_0$ the value of the scalar factor today.} $a_0=1$, we obtain
\begin{equation}
\label{Eqxi2}
\frac{\ud^2 \xi}{\ud a^2}  -\frac{3}{2} \frac{\nu/a}{\nu+a^3} \frac{\ud \xi}{\ud a}- 3 \frac{\nu/a^2}{\nu+a^3} \xi =0 \;.
 \end{equation}
 
\begin{figure}[htbp]
\begin{center}
\includegraphics[width=0.8\columnwidth]{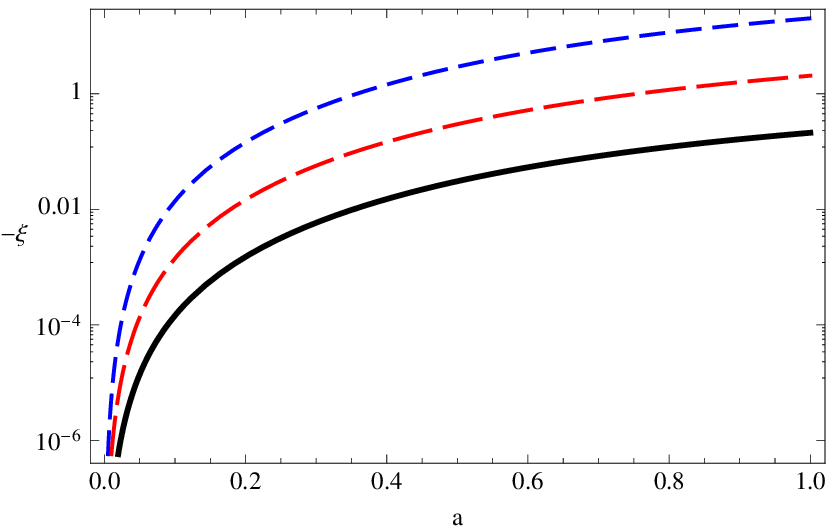}
\caption{Illustrative plot of $- \xi$,  as a function of  $a$, with $(\Omega_{\rm 0 \, m}/\Omega_{0 \, \Lambda})=3/7$. The lines, from short to long dashes, 
correspond to $\left(\ud \xi/\ud a\right)(a_{\rm rec})= - 10^{-5}, - 10^{-6}$ respectively; the black solid line corresponds to $\left(\ud \xi/\ud a\right)(a_{\rm rec})=-10^{-7}$.}
\label{fig:xi}
\end{center}
\end{figure}

\begin{figure}[htbp]
\begin{center}
\includegraphics[width=0.86\columnwidth]{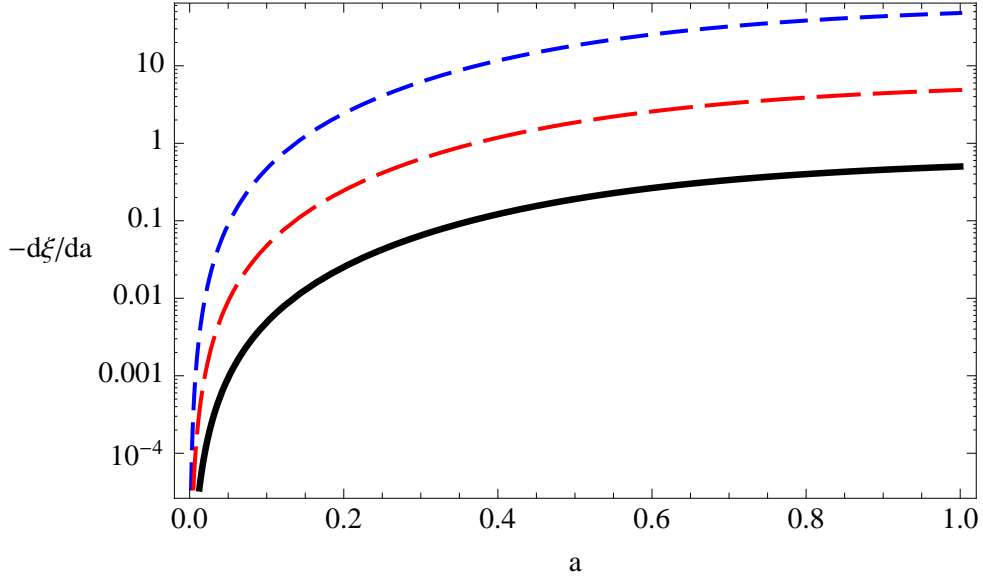}
\caption{Illustrative plot of $\left(- \ud \xi/\ud a\right)$,  as a function of  $a$, with $(\Omega_{0 \, \rm m}/\Omega_{0 \, \Lambda})=3/7$. The lines, from short to long dashes, 
correspond to $\left(\ud \xi/\ud a\right)(a_{\rm rec})= - 10^{-5}, - 10^{-6}$ respectively; the black solid line corresponds to $\left(\ud \xi/\ud a\right)(a_{\rm rec})=-10^{-7}$.}
\label{fig:derxi}
\end{center}
\end{figure}

In Figs.\ \ref{fig:xi} and \ref{fig:derxi}  we show the evolution of $(- \xi)$ and $\left(- \ud \xi/\ud a\right)$ as a function of  the scale-factor $a$. In particular we note that the value of $|\xi|$ grows rapidly after recombination and if the initial conditions are not sufficiently small, in future this effect could lead to an ``explosive phenomenon" 
for these models \cite{Arbuzova:2010iu}. It is obvious that in that regime our perturbative technique cannot be used.

Finally, it is useful to compare our formalism to the existing literature by considering the dimensionless quantity that quantifies the deviation from the $\Lambda$CDM reference model
introduced in Ref.\ \cite{Song:2006ej} (see also \cite{Zhang:2007ne, DeFelice:2010aj}) 
 \begin{equation}
 \label{B}
 B=m \frac{R'}{R} \frac{\mathcal{H}}{\left(\mathcal{H}' - \mathcal{H}^2\right)}\;,
 \end{equation}
where  \cite{Amendola:2006we, DeFelice:2010aj}
\begin{equation}
m=R \frac{(\partial^2 f/\partial R^2)}{ (\partial f/\partial R)}= \frac{(\ud V/ \ud \varphi)}{\varphi (\ud^2 V/ \ud \varphi^2)}\;.
 \end{equation}
In our case it becomes 
 \begin{equation}
 \label{Bxi}
B=\frac{\mathcal{H}^2}{\left(\mathcal{H}' - \mathcal{H}^2\right)} \left[a \frac{\ud \ln ( 1+\epsilon \xi)}{\ud a}\right] \;.
 \end{equation}
 We see immediately that, imposing $\xi,\xi'<0$ and $\epsilon >0$, we get the corresponding stability condition $B>0$ found in \cite{Song:2006ej}. 

\subsection{First-order perturbation equations}

In this section we analyse in detail the field perturbation equations through this iterative technique. From Eq.\ (\ref{pert-00}) we get
 \begin{eqnarray} 
 \label{pert-00_standard}
 6 \mathcal{H}\psi^{(1,0)}\,'  &-& \left(2 \nabla^2\psi^{(1,0)} +\frac{1}{3} \nabla^2\nabla^2 \chi^{(1,0)} \right) + 8 \pi a^2 \rho_{\rm m}^{(0)} \delta^{(1,0)} =0 \;, \\
  \nonumber \\
 6 \mathcal{H}\psi^{(1,1)}\,'  &-& \left(2 \nabla^2\psi^{(1,1)} +\frac{1}{3} \nabla^2\nabla^2 \chi^{(1,1)} \right) + 8 \pi a^2 \rho_{\rm m}^{(0)} \delta^{(1,1)} = - 3 \xi ' \psi^{(1,0)}\,' \nonumber \\
  \label{pert-00_xi}
 + 3 \mathcal{H} \xi^{(1,1)}\,' &-& \nabla^2   \xi^{(1,1)}  + \frac{a^2}{2} \left[V_0 - \frac{6}{a^2} \left(\mathcal{H}' + \mathcal{H}^2\right) \right]  \xi^{(1,1)}+  8 \pi a^2 \rho_{\rm m}^{(0)}  \left(\xi^{(1,1)} + \xi  \delta^{(1,0)}  \right) \;. \nonumber \\
  \end{eqnarray}
From Eq.\ (\ref{pert-0i}) we obtain
\begin{eqnarray} 
\label{pert-0i_standard}
2\psi^{(1,0)}\,' +\frac{1}{3} \nabla^2 \chi^{(1,0)}\,' &=& 0  \;, \\
 \nonumber \\
\label{pert-0i_xi}
2\psi^{(1,1)}\,' +\frac{1}{3} \nabla^2 \chi^{(1,1)}\,' &=&    \xi^{(1,1)}\,'  -  \mathcal{H} \xi^{(1,1)}  \;; 
\end{eqnarray}
while from Eq.\ (\ref{pert-ij})
\begin{eqnarray} 
\label{pert-ij_standard}
\chi^{(1,0)}\,''  + 2\mathcal{H}  \chi^{(1,0)}\,' + 2 \psi^{(1,0)} +\frac{1}{3} \nabla^2 \chi^{(1,0)} &=& 0 \;, \\
 \nonumber \\
\label{pert-ij_xi}
\chi^{(1,1)}\,''  + 2\mathcal{H}  \chi^{(1,1)}\,' + 2 \psi^{(1,1)} +\frac{1}{3} \nabla^2 \chi^{(1,1)} &=& 2  \xi^{(1,1)} -   \xi '  \chi^{(1,0)}\,'  \;.
\end{eqnarray}
Moreover, from Eq.\ (\ref{pert-trace}), we get
\begin{eqnarray}
\label{pert-trace_standard}
&& \frac{8 \pi a^2}{3}   \rho_{\rm m}^{(0)} \delta^{(1,0)}=\frac{2}{ \xi '} \left(\mathcal{H}'' -2\mathcal{H}^3 \right) \xi^{(1,1)} \;, \\
 \nonumber \\
&& \xi^{(1,1)}\,'' +2 \mathcal{H} \xi^{(1,1)}\,' - \nabla^2 \xi^{(1,1)}  - 3 \xi' \psi^{(1,0)}\,'  + \frac{2\xi}{ \xi '} \left(\mathcal{H}''  -2\mathcal{H}^3 \right) \xi^{(1,1)} - 2 \left(\mathcal{H}'+\mathcal{H}^2 \right) \xi^{(1,1)}\nonumber \\ \label{pert-trace_xi}
&& =\frac{8 \pi a^2}{3}   \rho_{\rm m}^{(0)} \delta^{(1,1)} \;;
\end{eqnarray} 
while, from Eq.\  (\ref{pert-eqmotion}),
\begin{eqnarray}
\label{pert-eqmotion_standard}
-6 \psi^{(1,0)}\,'' -18 \mathcal{H} \psi^{(1,0)}\,' + 4 \nabla^2 \psi^{(1,0)} + \frac{2}{3}  \nabla^2\nabla^2 \chi^{(1,0)} = \frac{6}{ \xi '} \left(\mathcal{H}'' -2\mathcal{H}^3 \right) \xi^{(1,1)}\;, \\
 \nonumber \\
\label{pert-eqmotion_xi}
-6 \psi^{(1,1)}\,'' -18 \mathcal{H} \psi^{(1,1)}\,' + 4 \nabla^2 \psi^{(1,1)} + \frac{2}{3}  \nabla^2\nabla^2 \chi^{(1,1)} = \frac{3}{ \xi '} \left(\mathcal{H}'' -2\mathcal{H}^3 \right) \xi^{(1,2)}\;.
\end{eqnarray} 
Let us note that in the LHS of Eqs.\ (\ref{pert-trace_standard}) and (\ref{pert-eqmotion_standard}) there is a term proportional to $ \xi^{(1,1)}$. 
This might appear as a mismatch of the $0$-th order and of the first order in $\epsilon$. This is not the case, however, since, in these equations, 
$\xi^{(1,1)}$ behaves as an auxiliary field which allows to connect these two equations. 
Indeed, replacing the RHS of Eq.\ (\ref{pert-trace_standard}) in 
Eq.\ (\ref{pert-eqmotion_standard}) one recovers the trace of the gravitational field equations of $\Lambda$CDM.

 Finally, from Eq.\ (\ref{pert-conteq}), we find
 \begin{eqnarray}
 \label{pert-conteq_standard}
\delta^{(1,0)}\,'=3\psi^{(1,0)}\,' \;, \\
 \nonumber \\
 \label{pert-conteq_xi}
\delta^{(1,1)}\,'=3\psi^{(1,1)}\,' \;.
\end{eqnarray}

At this point, from Eqs.\ (\ref{pert-00_standard}),   (\ref{pert-0i_standard}),  (\ref{pert-ij_standard}),  (\ref{pert-eqmotion_standard}) [substituting the RHS of Eq.\ (\ref{pert-trace_standard}) in the RHS of Eq.\  (\ref{pert-eqmotion_standard})] and (\ref{pert-conteq_standard}), we obtain the perturbation equations in the synchronous gauge in general relativity (see Appendix \ref{A} and Refs.\ \cite{Ma:1995ey, Matarrese:1997ay}). Instead, from the other equations we are able to get 
their correction terms, i.e. $ \delta^{(1,1)}$, $2 \nabla^2\psi^{(1,1)} +(1/3) \nabla^2 \nabla^2 \chi^{(1,1)}$. In particular, knowing that $\mathcal{H}''=\mathcal{H}'\mathcal{H}+\mathcal{H}^3$, and after some tedious calculations,
we derive the following relations\footnote{In order to illustrate better that the background is the same of the $\Lambda$CDM model for $\epsilon\,\xi \ll 1$, let us consider for example Eq.\ (\ref{delta11}). 
In the RHS of this equation most of the coefficients are proportional to $\mathcal{H}^2$ or $\mathcal{H}'/\mathcal{H}^2$. In this case, they 
describe the same background as the $\Lambda$CDM model because the 
corrections are of order $\epsilon^2$. In other words, they are negligible 
at first order in $\epsilon$.}    
\begin{eqnarray}
\label{delta11}
\frac{8 \pi a^2}{3} \rho_{\rm m}^{(0)} \delta^{(1,1)} &=& \frac{1}{3} \left[ \left( 7\mathcal{H}^2 - 9 \mathcal{H}' + 2 \frac{\mathcal{H}'^2}{\mathcal{H}^2}+ 4\pi a^2   \rho_{\rm m}^{(0)} \right)  \left(- a \frac{\ud \xi}{\ud a} \right)  \right. \nonumber \\
&+& \left. 2 \left( 5 \mathcal{H}' - 3 \mathcal{H}^2 -2\frac{\mathcal{H}'^2}{\mathcal{H}^2} + 4 \pi a^2 \rho_{\rm m}^{(0)}  \right) \xi \right] \delta^{(1,0)}  \nonumber \\
&+&  \frac{1}{3} \left[2\left(  4 \mathcal{H} - \frac{\mathcal{H}'}{\mathcal{H}} \right)\left(- a \frac{\ud \xi}{\ud a} \right) + 4\left(-\mathcal{H} + \frac{\mathcal{H}'}{\mathcal{H}}\right) \xi \right] \delta^{(1,0)}\,' - \frac{1}{3} \left(- a \frac{\ud \xi}{\ud a} \right) \nabla^2\delta^{(1,0)} \;, \nonumber \\
\end{eqnarray}
\begin{eqnarray}
\label{psi11}
2 \nabla^2\psi^{(1,1)} &+&\frac{1}{3} \nabla^2 \nabla^2 \chi^{(1,1)}= 
\left\{\frac{\mathcal{H}^2}{4 \pi a^2 \rho_{\rm m}^{(0)}}\left[\left(18 \mathcal{H}^2 - 35 \mathcal{H}' +23 \frac{\mathcal{H}'^2}{\mathcal{H}^2}-6 \frac{\mathcal{H}'^3}{\mathcal{H}^4} \right)\left(- a \frac{\ud \xi}{\ud a}\right)   \right.   \right.  \nonumber \\
&+&  \left.   \left. 2 \left(16  \mathcal{H}'  - 5  \mathcal{H}^2 - 17 \frac{\mathcal{H}'^2}{\mathcal{H}^2} + 12 \frac{\mathcal{H}'^3}{\mathcal{H}^4}  \right) \xi\right]   + \left( 13  \mathcal{H}^2 - 10 \mathcal{H}'  +2 \frac{\mathcal{H}'^2}{\mathcal{H}^2}\right) \left(- a \frac{\ud \xi}{\ud a}\right)  \right.   \nonumber \\
&-& \left. 2 \left(5 \mathcal{H}^2 - 7\mathcal{H}' +2 \frac{\mathcal{H}'^2}{\mathcal{H}^2} \right) \xi + 4 \pi a^2 \rho_{\rm m}^{(0)} \left(- a \frac{\ud \xi}{\ud a}\right) \right\}  \delta^{(1,0)}   \nonumber \\
&+& \left\{\frac{\mathcal{H}}{4 \pi a^2 \rho_{\rm m}^{(0)}}\left[\left(25 \mathcal{H}^2 - 19\mathcal{H}' + 6 \frac{\mathcal{H}'^2}{\mathcal{H}^2} \right) \left(- a \frac{\ud \xi}{\ud a}\right)  + 3 \left(10\mathcal{H}'  -6\mathcal{H}^2 -4  \frac{\mathcal{H}'^2}{\mathcal{H}^2}\right)\xi \right]   \right.   \nonumber \\
&+ &\left. \left(7 \mathcal{H} - 2 \frac{\mathcal{H}'}{\mathcal{H}} \right) \left(- a \frac{\ud \xi}{\ud a}\right)  + 2 \left(- \mathcal{H}  +2  \frac{\mathcal{H}'}{\mathcal{H}}\right)\xi  \right\}  \delta^{(1,0)} \,'   \nonumber \\
&+& \left\{  \frac{1}{4 \pi a^2 \rho_{\rm m}^{(0)}} \left[ \left(-3 \mathcal{H}^2 + \mathcal{H}'  \right)\left(- a \frac{\ud \xi}{\ud a}\right) +  2 \left(\mathcal{H}^2 - \mathcal{H}'   \right) \xi \right]  - \frac{2}{3} \left(- a \frac{\ud \xi}{\ud a}\right) \right\}  \nabla^2\delta^{(1,0)}   \nonumber \\
&-& \frac{\mathcal{H}}{4 \pi a^2 \rho_{\rm m}^{(0)}}\left(- a \frac{\ud \xi}{\ud a}\right)   \nabla^2\delta^{(1,0)}\,' \;.
\end{eqnarray}
Moreover, by linearizing the solution of the continuity equation~(\ref{pert-conteq_xi}) we obtain
\begin{equation}
\psi^{(1,1)}(\eta,\mbox{\bf x})= \psi^{(1,1)}_0 (\mbox{\bf x})+\frac{1}{3}\left(\delta^{(1,1)}(\eta,\mbox{\bf x})-\delta^{(1,1)}_0 (\mbox{\bf x}) \right)\;.
\end{equation}
We denote by a subscript $0$ the condition at the present time of the referred quantity. Then
\begin{eqnarray}
\label{chi11}
\frac{1}{3} \nabla^2 \nabla^2 \chi^{(1,1)} &=& 
\left\{\frac{\mathcal{H}^2}{4 \pi a^2 \rho_{\rm m}^{(0)}}\left[\left(18 \mathcal{H}^2 - 35 \mathcal{H}' +23 \frac{\mathcal{H}'^2}{\mathcal{H}^2}-6 
\frac{\mathcal{H}'^3}{\mathcal{H}^4} \right)\left(- a \frac{\ud \xi}{\ud a}\right)   \right.   \right.  \nonumber \\
&+&  \left.   \left. 2 \left(16  \mathcal{H}'  - 5  \mathcal{H}^2 - 17 \frac{\mathcal{H}'^2}{\mathcal{H}^2} + 12 \frac{\mathcal{H}'^3}{\mathcal{H}^4}  
\right) \xi\right]   + \left( 13  \mathcal{H}^2 - 10 \mathcal{H}'  +2 \frac{\mathcal{H}'^2}{\mathcal{H}^2}\right) \left(- a \frac{\ud \xi}{\ud a}\right)  \right.  
 \nonumber \\
&-& \left. 2 \left(5 \mathcal{H}^2 - 7\mathcal{H}' +2 \frac{\mathcal{H}'^2}{\mathcal{H}^2} \right) \xi + 4 \pi a^2 \rho_{\rm m}^{(0)} \left(- a \frac{\ud \xi}{\ud a}\right) \right\}  \delta^{(1,0)}   \nonumber \\
&+& \left\{\frac{\mathcal{H}}{4 \pi a^2 \rho_{\rm m}^{(0)}}\left[\left(25 \mathcal{H}^2 - 19\mathcal{H}' + 6 \frac{\mathcal{H}'^2}{\mathcal{H}^2} \right) \left(- a \frac{\ud \xi}{\ud a}\right)  + 3 \left(10\mathcal{H}'  
-6\mathcal{H}^2 -4  \frac{\mathcal{H}'^2}{\mathcal{H}^2}\right)\xi \right]   \right.   \nonumber \\
&+ &\left. \left(7 \mathcal{H} - 2 \frac{\mathcal{H}'}{\mathcal{H}} \right) \left(- a \frac{\ud \xi}{\ud a}\right)  + 2 \left(- \mathcal{H}  +2  \frac{\mathcal{H}'}{\mathcal{H}}\right)\xi  \right\}  \delta^{(1,0)} \,'   \nonumber \\
&+& \left\{ \frac{1}{12 \pi a^2 \rho_{\rm m}^{(0)}} \left[ 2\left(-8 \mathcal{H}^2 + 6 \mathcal{H}' -  \frac{\mathcal{H}'^2}{\mathcal{H}^2}\right)\left(- a \frac{\ud \xi}{\ud a}\right)  + 4 \left(3 \mathcal{H}^2 -4 \mathcal{H}' 
 +  \frac{\mathcal{H}'^2}{\mathcal{H}^2}\right)\xi  \right]  \right.   \nonumber \\
&-& \left. \left(- a \frac{\ud \xi}{\ud a}  \right)-3 \xi\right\}  \nabla^2\delta^{(1,0)}   \nonumber \\
&+& \frac{1}{12 \pi a^2 \rho_{\rm m}^{(0)}} \left[\left(-11 \mathcal{H} + 2\frac{\mathcal{H}'}{\mathcal{H}}\right)\left(- a \frac{\ud \xi}{\ud a}\right)  +4 \left( \mathcal{H} - 
\frac{\mathcal{H}'}{\mathcal{H}}\right)\xi \right]  \nabla^2\delta^{(1,0)}\,'     \nonumber \\
&+&  \frac{1}{12 \pi a^2 \rho_{\rm m}^{(0)}}  \left(- a \frac{\ud \xi}{\ud a}  \right) \nabla^2 \nabla^2\delta^{(1,0)} -2 \nabla^2\left(\psi^{(1,1)}_0 -\frac{1}{3}\delta^{(1,1)}_0 \right) \;.
\end{eqnarray}
At this point, one can remove the residual gauge ambiguity of the synchronous coordinates by imposing that\footnote{In addition, we can conclude 
that $\delta_0^{(1)}=-(1/2)\nabla^2 \chi^{(1)}_0$, see Appendix \ref{A}.} $\delta_0^{(1,1)}=-(1/2)\nabla^2 \chi^{(1,1)}_0$. Therefore, evaluating Eq.\  (\ref{chi11}) to present time, we can determine immediately the value of $\psi^{(1,1)}_0$.

Eqs. (\ref{delta11}), (\ref{psi11}) and (\ref{chi11}), which govern the lowest-order modified gravity corrections w.r.t. $\Lambda$CDM in the behavior of scalar perturbations, represent the main result of this paper. Let us  stress two important facts: 1) all our expressions are completely determined by using the well-known results of the $\Lambda$CDM model (see Appendix \ref{A}) and the dynamical solution of $\xi$; 2) through this method it is possible to obtain analytically a more precise result if we consider the next orders in $\epsilon$.
 
 In the next section we will analyze in detail the evolution of perturbations in the Poisson gauge.

\section{From the synchronous to the Poisson gauge} \label{poissongauge}

In this section we obtain the evolution of cosmological perturbations in the Poisson gauge\footnote{In the Poisson gauge one scalar degree of freedom is eliminated 
from the $g_{0i}$ component of the metric, and one scalar and two vector degrees of freedom are eliminated from $g_{ij}$.} (also known as the conformal Newtonian gauge or the 
longitudinal gauge) through a gauge transformation of the results obtained in the synchronous gauge in the previous section.
In particular we will follow the approach used in Ref.\ \cite{Matarrese:1997ay} (e.g. see also \cite{Ma:1995ey}). 
Setting linear vector and tensor modes to zero, the flat linear metric becomes $ds^2=a^2 [- (1+ 2 \Phi^{(1)}_{\rm p})d\eta^2 + (1- 2 \Psi^{(1)}_{\rm p}) \delta_{ij} dx^i dx^j ]$. 
Instead, perturbing the mass-density and fluid four-velocity we get $\rho_{\rm m} =\rho_{\rm m}^{(0)} (1 + \delta_{\rm p}^{(1)})$ and $u^{\mu}=(\delta_0^\mu+v^{(1) \, \mu}_{\rm p})/a$, 
where $v^{(1) \, 0}=-\Phi^{(1)}_{\rm p}$  and $v^{(1) \, i}_{\rm p}=\partial^i v_{\rm p}^{(1)}$.

Considering the perturbations at the same space-time coordinate values, the synchronous gauge and the conformal Newtonian gauge can be related by the following 
relations \cite{Matarrese:1997ay, Ma:1995ey}
\begin{eqnarray} \label{Phi_p}
-2 \Phi^{(1)}_{\rm p} &=&  \chi^{(1)}\,'' +  \mathcal{H}  \chi^{(1)}\,' \;, \\
\label{Psi_p}
2 \Psi^{(1)}_{\rm p} &=& 2 \psi^{(1)} + \frac{1}{3} \nabla^2 \chi^{(1)} +  \mathcal{H}  \chi^{(1)}\,' \;, \\
\label{delta_p}
\delta_{\rm p}^{(1)} &=& \delta^{(1)} + \frac{3}{2}   \mathcal{H} \chi^{(1)}\,' \;, \\
\label{v_p}
v_{\rm p}^{(1)} &=& \frac{1}{2}  \chi^{(1)}\,' \;.
\end{eqnarray}

At this point, as we have already done in section \ref{Technique}, we can split $ \Phi^{(1)}_{\rm p}$,  $\Psi^{(1)}_{\rm p}$,  $\delta_{\rm p}^{(1)}$ and $v_{\rm p}^{(1)}$ in the following way
\begin{eqnarray}
\Phi^{(1)}_{\rm p} &=& \Phi^{(1,0)}_{\rm p}  + \epsilon \, \Phi^{(1,1)}_{\rm p}  + ... \;, \\
\Psi^{(1)}_{\rm p} &=& \Psi^{(1,0)}_{\rm p} + \epsilon \, \Psi^{(1,1)}_{\rm p}  + ...  \;, \\
\delta_{\rm p}^{(1)} &=&  \delta_{\rm p}^{(1,0)}  + \epsilon \,  \delta_{\rm p}^{(1,1)}  + ... \;, \\
v_{\rm p}^{(1)} &=&  v_{\rm p}^{(1,0)}  + \epsilon \, v_{\rm p}^{(1,1)}   + ... \;.
\end{eqnarray}

Also in this case the terms  $\Phi^{(1,0)}_{\rm p}$,  $\Psi^{(1,0)}_{\rm p}$,  $\delta_{\rm p}^{(1,0)}$ and $v_{\rm p}^{(1,0)}$  are the perturbation terms that one obtains in a $\Lambda$CDM model.
Now, we want to study and determine the terms to order $\epsilon$. 
Immediately we note that using Eqs.\ (\ref{pert-0i_xi}), and (\ref{pert-conteq_xi}) we can obtain $v_{\rm p}^{(1,1)}$ and, consequently, $\delta_{\rm p}^{(1,1)}$. Indeed
\begin{eqnarray}\label{v_p11}
v_{\rm p}^{(1,1)} &=&  \frac{1 }{8 \pi a^2 \rho_{\rm m}^{(0)}} \left\{  \left[\left( -18 \mathcal{H}^3 + 35 \mathcal{H}' \mathcal{H} - 23 \frac{\mathcal{H}'^2}{\mathcal{H}} + 6 
\frac{\mathcal{H}'^3}{\mathcal{H}^3} \right)\left(- a \frac{\ud \xi}{\ud a}\right)  \right.   \right. \nonumber \\
&-&  \left.   \left. 2     \left(16  \mathcal{H}' \mathcal{H} - 5  \mathcal{H}^3 - 17 \frac{\mathcal{H}'^2}{\mathcal{H}} + 6 \frac{\mathcal{H}'^3}{\mathcal{H}^3}  \right) \xi  
  -                       4    \pi a^2 \rho_{\rm m}^{(0)} \left(7\mathcal{H} - 2 \frac{\mathcal{H}'}{\mathcal{H}} \right) \left(- a \frac{\ud \xi}{\ud a}\right)    \right.   \right. \nonumber \\
&+& \left.   \left. 16 \pi a^2 \rho_{\rm m}^{(0)}  \left(\mathcal{H} - \frac{\mathcal{H}'}{\mathcal{H}} \right) \xi \right]   \nabla^{-2} \delta^{(1,0)}  
  -     \left[    \left(25 \mathcal{H}^2 - 19 \mathcal{H}' + 6 \frac{\mathcal{H}'^2}{\mathcal{H}^2} \right) \left(- a \frac{\ud \xi}{\ud a}\right)  \right.  \right. \nonumber \\
&+& \left.   \left.  6 \left(5 \mathcal{H}' -3 \mathcal{H}^2- 2 \frac{\mathcal{H}'^2}{\mathcal{H}^2}\right)\xi  
  +                        8   \pi a^2 \rho_{\rm m}^{(0)}  \xi \right]   \nabla^{-2}\delta^{(1,0)}\,'      \right.  \nonumber \\
&+&  \left.  \left[  \left(3 \mathcal{H} - \frac{\mathcal{H}'}{\mathcal{H}}\right)\left(- a \frac{\ud \xi}{\ud a}\right)  + 2 \left(- \mathcal{H}+ \frac{\mathcal{H}'}{\mathcal{H}}\right)\xi \right] \delta^{(1,0)}
   +    \left(- a \frac{\ud \xi}{\ud a}\right) \delta^{(1,0)}\,'    \right\} \;;
\end{eqnarray}
where  $\nabla^{-2}$ stands for  the inverse of the Laplacian operator; while from Eqs.\ (\ref{delta_p}) and (\ref{delta11}) we get
\begin{eqnarray}\label{v_p11}
\delta_{\rm p}^{(1,1)} &=& \frac{1}{4 \pi a^2 \rho_{\rm m}^{(0)}} \left[ \left( 8 \mathcal{H}^2 - 6 \mathcal{H}' +  \frac{\mathcal{H}'^2}{\mathcal{H}^2} \right)  \left(- a \frac{\ud \xi}{\ud a} \right)  
   +  \left( 8 \mathcal{H}' - 6 \mathcal{H}^2 -2\frac{\mathcal{H}'^2}{\mathcal{H}^2}  \right) \xi  \right. \nonumber \\
&+& \left.  2 \pi a^2   \rho_{\rm m}^{(0)} \left(- a \frac{\ud \xi}{\ud a} \right)  + 4 \pi a^2 \rho_{\rm m}^{(0)}   \xi\right] \delta^{(1,0)} 
+ \frac{1}{8 \pi a^2 \rho_{\rm m}^{(0)}}  \left[ \left(  11 \mathcal{H} - 2 \frac{\mathcal{H}'}{\mathcal{H}} \right)\left(- a \frac{\ud \xi}{\ud a} \right)  \right. \nonumber \\
&+&  \left. 4\left(-\mathcal{H} + \frac{\mathcal{H}'}{\mathcal{H}}\right) \xi \right] \delta^{(1,0)}\,' - \frac{1}{8 \pi a^2 \rho_{\rm m}^{(0)}} \left(- a \frac{\ud \xi}{\ud a} \right) \nabla^2\delta^{(1,0)}  \nonumber \\
&+&  \frac{3 \mathcal{H}^2}{8 \pi a^2 \rho_{\rm m}^{(0)}}  \left[ \left( -18 \mathcal{H}^2 + 35 \mathcal{H}'  - 23 \frac{\mathcal{H}'^2}{\mathcal{H}^2} + 6 \frac{\mathcal{H}'^3}{\mathcal{H}^4} \right) \left(- a \frac{\ud \xi}{\ud a}\right)   \right. \nonumber \\
&-&   \left. 2     \left(16  \mathcal{H}' \mathcal{H} - 5  \mathcal{H}^3 - 17 \frac{\mathcal{H}'^2}{\mathcal{H}} + 6 \frac{\mathcal{H}'^3}{\mathcal{H}^3}  \right) \xi  
  -      4    \pi a^2 \rho_{\rm m}^{(0)} \left(7\mathcal{H} - 2 \frac{\mathcal{H}'}{\mathcal{H}} \right) \left(- a \frac{\ud \xi}{\ud a}\right)     \right. \nonumber \\
&+&  \left. 16 \pi a^2 \rho_{\rm m}^{(0)}  \left(\mathcal{H} - \frac{\mathcal{H}'}{\mathcal{H}} \right) \xi \right]   \nabla^{-2} \delta^{(1,0)} - \frac{3 \mathcal{H}}{8 \pi a^2 \rho_{\rm m}^{(0)}}  \left[  \left(25 \mathcal{H}^2 - 19 \mathcal{H}' + 6 \frac{\mathcal{H}'^2}{\mathcal{H}^2} \right) \left(- a \frac{\ud \xi}{\ud a}\right)   \right. \nonumber \\
&+&  \left.  3 \left(10 \mathcal{H}' -6 \mathcal{H}^2- 4 \frac{\mathcal{H}'^2}{\mathcal{H}^2}\right)\xi  
  +      8   \pi a^2 \rho_{\rm m}^{(0)}  \xi \right]   \nabla^{-2}\delta^{(1,0)}\,'    \;.
\end{eqnarray}

Instead from Eqs.\  (\ref{Phi_p}), (\ref{v_p}),  (\ref{v_p11}), (\ref{pert-ij_xi}) and (\ref{psi11}) we find
\begin{eqnarray}\label{phi_p11}
\Phi^{(1,1)}_{\rm p}  &=&  \left[ \left(3 \mathcal{H}^2  - 4 \mathcal{H}' + \frac{\mathcal{H}'^2}{\mathcal{H}^2} \right)  \left(- a \frac{\ud \xi}{\ud a}\right) -   \left( 3 \mathcal{H}^2  -5 \mathcal{H}' + 2 \frac{\mathcal{H}'^2}{\mathcal{H}^2} \right)  \xi  \right. \nonumber \\
&+& \left.  2 \pi a^2 \rho_{\rm m}^{(0)}  \left(- a \frac{\ud \xi}{\ud a}\right) \right]  \nabla^{-2} \delta^{(1,0)} 
  +    \frac{1}{2}  \left[ \left(9 \mathcal{H} - 2\frac{\mathcal{H}'}{\mathcal{H}}\right)\left(- a \frac{\ud \xi}{\ud a}\right)    \right. \nonumber \\
&+& \left. 4 \left(- \mathcal{H} + \frac{\mathcal{H}'}{\mathcal{H}}\right)\xi \right]  \nabla^{-2}\delta^{(1,0)}\,' - \frac{4}{3} \left(- a \frac{\ud \xi}{\ud a}\right) \delta^{(1,0)}  \;. 
\end{eqnarray}
Moreover from Eqs.\  (\ref{Psi_p}), (\ref{v_p}),  (\ref{v_p11}) and (\ref{psi11}) we obtain
\begin{eqnarray}\label{psi_p11}
\Psi^{(1,1)}_{\rm p}  &=& \left[ \left(3 \mathcal{H}^2  - 4 \mathcal{H}' + \frac{\mathcal{H}'^2}{\mathcal{H}^2} \right)  \left(- a \frac{\ud \xi}{\ud a}\right) -   \left( 3 \mathcal{H}^2  -5 \mathcal{H}' + 2 \frac{\mathcal{H}'^2}{\mathcal{H}^2} \right)  \xi  \right. \nonumber \\
&+& \left.  2 \pi a^2 \rho_{\rm m}^{(0)}  \left(- a \frac{\ud \xi}{\ud a}\right) \right]  \nabla^{-2} \delta^{(1,0)} 
  +    \frac{1}{2}  \left[ \left(7 \mathcal{H} - 2\frac{\mathcal{H}'}{\mathcal{H}}\right)\left(- a \frac{\ud \xi}{\ud a}\right)    \right. \nonumber \\
&+& \left. 4 \left(- \mathcal{H} + \frac{\mathcal{H}'}{\mathcal{H}}\right)\xi \right]  \nabla^{-2}\delta^{(1,0)}\,' - \frac{2}{3} \left(- a \frac{\ud \xi}{\ud a}\right) \delta^{(1,0)}  \;. 
\end{eqnarray}

Finally, another useful quantity is the anisotropic contribution $\Pi^{(1)}$ which is one of the parameters that allow to quantify the departure of $f(R)$ gravity from the standard  $\Lambda$CDM model \cite{Pogosian:2007sw}. Indeed in our formalism $\Pi^{(1)}=\Psi_{\rm p}^{(1)} - \Phi_{\rm p}^{(1)} = \epsilon (\Psi^{(1,1)}_{\rm p} - \Phi^{(1,1)}_{\rm p})$ because $\Psi^{(1,0)}_{\rm p} = \Phi^{(1,0)}_{\rm p}$ (see appendix \ref{B}).
Then
\begin{eqnarray}\label{psi-phi}
\Pi^{(1)}/\epsilon=2 \left(- a \frac{\ud \xi}{\ud a}\right)  \left(\frac{1}{3} \delta^{(1,0)} -  \mathcal{H} \nabla^{-2} \delta^{(1,0)}\,' \right) \;.
\end{eqnarray}

\section{Comparison with the ``quasi-static" approximation} \label{quasi-static}

As it is well known, when we consider scales deep inside the Hubble radius, in order to derive 
the equation of matter perturbations approximately, one uses the quasi-static approximation (for details, for example, see \cite{DeFelice:2010aj}). 
This section is devoted  to recover this approximation, as an important example of how our formalism confront with and recover some known results in the literature. 
In particular we want to calculate the Poisson equation for these models with the approach studied in this work.
Defining $\Phi_{\rm eff}^{(1)}=-(\Phi_{\rm p}^{(1)}+\Psi_{\rm p}^{(1)})/2$   \cite{DeFelice:2010aj},  from Eqs.\ (\ref{Phi_p}) and (\ref{Psi_p}), we get 
\begin{eqnarray}\label{phieff1}
 \Phi_{\rm eff}^{(1)}=-\frac{1}{4} \left( 2 \psi^{(1)} +\frac{1}{3} \nabla^2 \chi^{(1)} - \psi^{(1)}\,'' \right) \;.
\end{eqnarray}
Then, knowing that  $\Phi_{\rm eff}^{(1)}=\Phi_{\rm eff}^{(1,0)}+ \epsilon \Phi_{\rm eff}^{(1,1)}$ and using Eqs.\ (\ref{pert-ij_xi}),  (\ref{pert-00_xi}) and (\ref{pert-0i_xi}), we obtain 
\begin{eqnarray}\label{phieff11}
\nabla^2 \Phi_{\rm eff}^{(1,1)}=-\frac{1}{2} \left[ 3( \mathcal{H}'- \mathcal{H}^2)\xi^{(1,1)} +8\pi a^2 \rho_{\rm m}^{(0)} \left(\delta^{(1,1)} - \xi \delta^{(1,0)}   \right) \right]\;.
\end{eqnarray}
At this point, let us consider in detail Eq.\ (\ref{pert-trace_xi}).
In that scales we can drop the terms with temporal derivatives when we compare them with spatial gradients term in $\xi^{(1,1)}$. Then we find
\begin{equation}\label{xi11qa}
\left[  \frac{2\xi}{ \xi '} \left(\mathcal{H}''  -2\mathcal{H}^3 \right) - 2 \left(\mathcal{H}'+\mathcal{H}^2 \right)- \nabla^2\right] \xi^{(1,1)} 
=\frac{1}{3} 8 \pi a^2  \rho_{\rm m}^{(0)} \delta^{(1,1)}   -  \xi' \delta^{(1,0)}\,'\;.
\end{equation}
As we see from Figs.\  \ref{fig:xi}  and \ref{fig:derxi} ,  $[\xi/ (\ud \xi/ \ud a )]$ is less of $1/10$. Moreover, considering scales where the square of wavenumber $k$ is larger than $ \mathcal{H}'$  and $\mathcal{H}^2$, we conclude that 
$- \nabla^2 \xi^{(1,1)} \simeq [(8 \pi a^2/3)  \rho_{\rm m}^{(0)} \delta^{(1,1)}   -  \xi' \delta^{(1,0)}\,']$. Consequently, in this case, the additive term proportional to $\xi^{(1,1)}$ becomes negligible with respect to the other terms in Eq.\ (\ref{phieff11}). Then
\begin{eqnarray}\label{phieff11_2}
\nabla^2 \Phi_{\rm eff}^{(1,1)} \simeq -4\pi a^2 \rho_{\rm m}^{(0)} \left(\delta^{(1,1)} - \xi \delta^{(1,0)}   \right) \;.
\end{eqnarray}
 
 On the other hand, starting from the literature (for example, see Refs.\  \cite{Song:2006ej, DeFelice:2010aj}) and assuming the quasi-static approximation (i.e. when $\nabla^2 |X| \gg \mathcal{H}^2 |X|$ and $|X'|  <  \mathcal{H}|X| $, where\footnote{In general, for these models, one also adds the approximation that corrisponds to $|\varphi (\ud^2 V/\ud \varphi^2)| \gg  |(\ud V/\ud \varphi)|$. Let us stress that this condition is automatically satisfied through our approach (see section \ref{FLRW}).} $X=\Phi_{\rm p}^{(1)},\Psi_{\rm p}^{(1)}, \varphi, \varphi', \phi^{(1)}, \phi^{(1)}\,'$), we find
$\nabla^2 \Phi_{\rm eff}^{(1)}=-4\pi a^2 \rho_{\rm m}^{(0)}\delta^{(1)}/ \varphi$.
Then we can  quickly obtain the same result. Indeed  
\begin{eqnarray}
\nabla^2 \Phi_{\rm eff}^{(1)} \simeq - 4\pi a^2 \rho_{\rm m}^{(0)}\left[\delta^{(1,0)} + \epsilon  \left(\delta^{(1,1)} - \xi \delta^{(1,0)} \right) \right]\;.
\end{eqnarray} 
However, let us stress that our formalism is designed in order to easily account for effects that go beyond the 
quasi-static approximation, in particular those taking place  on scales comparable to the horizon size, where time derivatives 
cannot be neglected. Indeed they could be important, for example, for 
accurate calculation of the Integrated Sachs-Wolfe (ISW) 
effect and the large-scale matter power spectrum (for example, see \cite{Zuntz:2011aq}).

\section{Matter Power Spectrum} \label{MPS}

In this section we compute the matter power spectrum $ P(k,a)$:
\begin{equation}
\langle \delta^{(1)}({\bf k}, a)\; \delta^{(1)}({\bf k}', a) \rangle
= (2\pi)^3 \delta^3_D ({\bf k}+ {\bf k}') P(k,a)\;,
\end{equation}
where $k=|{\bf k}|$. Within our approach, it becomes
 \begin{equation}
 P(k,a)=P^{(0)}(k,a)+\epsilon P^{(1)}(k,a)
  \end{equation}
where  $P^{(0)}(k,a)$ is the matter power spectrum in the $\Lambda$CDM model. 
In terms of the primordial power spectrum and the transfer
function, we have
\begin{equation}
P^{(0)}(k,a)=P_{\rm prim}(k)T^2(k)\left({D(a) \over
D(0)}\right)^2\, ,
\end{equation}
 where $D $ is the growing mode of $\delta^{(1,0)}$, see Appendix \ref{A}.
 From Eq.\ (\ref{delta11}) we get
  \begin{equation}
 P^{(1)}(k)=2\mathcal{A}(k,a)P^{(0)}(k,a)
  \end{equation}
 with
\begin{eqnarray}
 3\Omega_m(a) \mathcal{A}(k,a)&=& \left[7-\frac{15}{2}\Omega_{\rm m}(a)+\frac{9}{2}\Omega_\Lambda(a)\right]+2\left[\Omega_{\rm m}(a)-\frac{1}{2}\Omega_\Lambda(a)\right]^2 \left(- a \frac{\ud \xi}{\ud a}\right)\nonumber \\
 &+&2\left\{-3+\frac{13}{2}\Omega_{\rm m}(a)-\frac{5}{2}\Omega_\Lambda(a)-2\left[\Omega_{\rm m}(a)-\frac{1}{2}\Omega_\Lambda(a)\right]^2\right\}\xi(a)\nonumber \\
 &+& f(a) \bigg\{2\left[4-\Omega_{\rm m}(a)+\frac{1}{2}\Omega_\Lambda(a)\right] \left(- a \frac{\ud \xi}{\ud a}\right)-4\left[1-\Omega_{\rm m}(a)+\frac{1}{2}\Omega_\Lambda(a)\right]\xi(a)\bigg\}\nonumber \\
 &+& \left(- a \frac{\ud \xi}{\ud a}\right)\left(\frac{k}{\mathcal{H}}\right)^2  \;,
 \end{eqnarray}
where $f$ is the growth rate, $\Omega_{\rm m}(a)$ is the density parameter of non-relativistic matter and $\Omega_\Lambda(a)$ is  the density parameter of the cosmological constant in a $\Lambda$CDM model, see Appendixes \ref{A} and \ref{B}.

Then, defining the initial solutions as in Section \ref{FLRW}, we can finally derive  $P(k,a=1)$. For simplicity, we fix $\xi(a_{\rm rec})=0$ and $\left(\ud \xi/\ud a\right)(a_{\rm rec})=-5.\;10^{-9}$ in order to get $\xi(a=1) \simeq -10^{-2}$ and  $\left(\ud \xi/\ud a\right)(a=1) \simeq-2.43 \cdot 10^{-2}$.
\begin{figure}[htbp]
\begin{center}
\includegraphics[width=0.46\columnwidth]{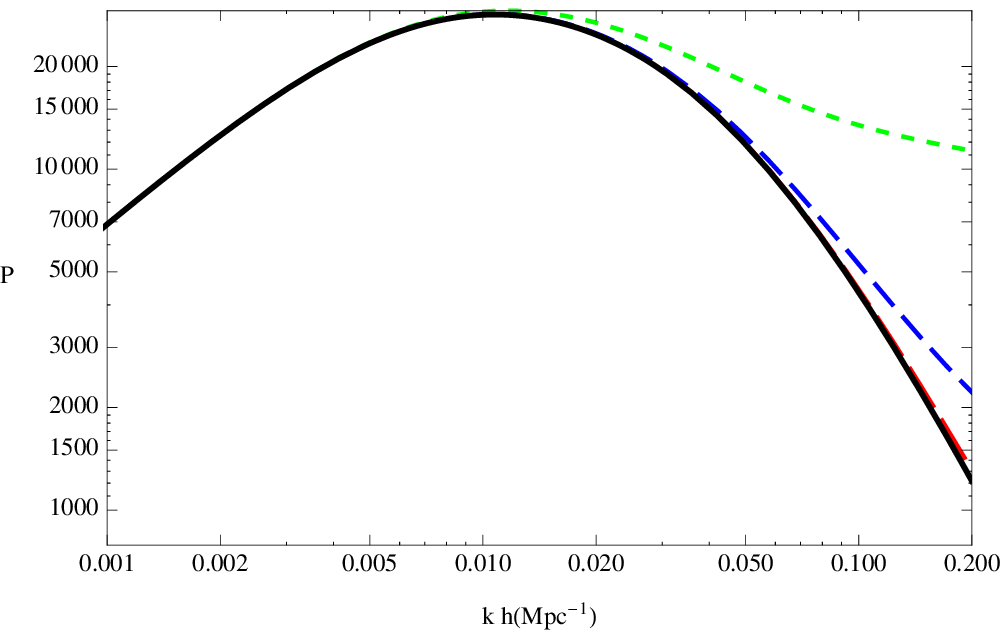}
\includegraphics[width=0.48\columnwidth]{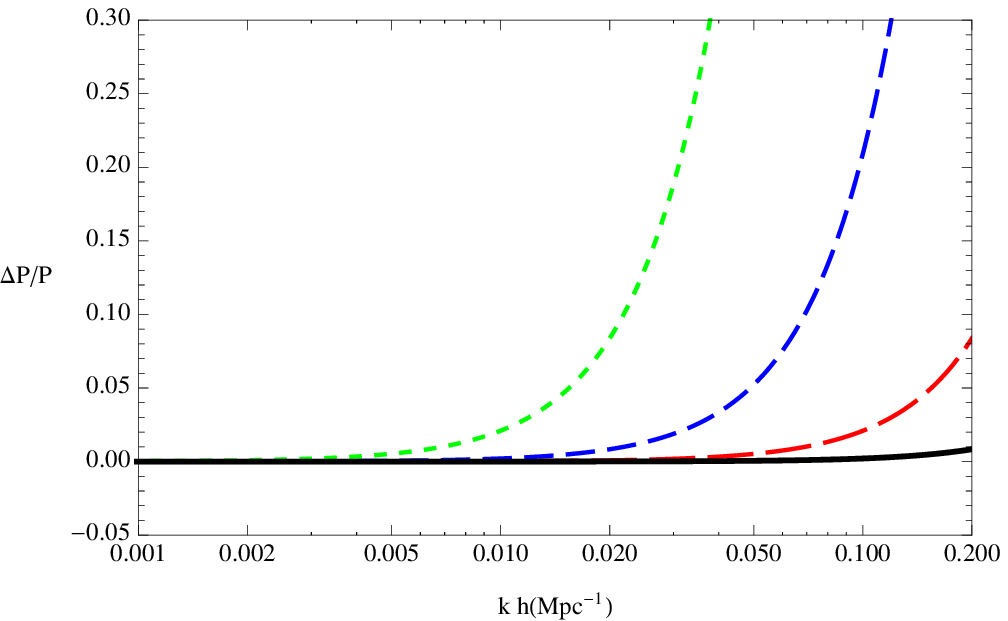}
\caption{Illustrative plot of $P$ and $\left({\Delta P / P}\right)$,  as a function of  $k$ and for $a=1$. The lines, from short to long dashes, 
correspond to $\epsilon=  10^{-4},  10^{-5},  10^{-6}$ respectively; the black solid line corresponds to $\epsilon=10^{-7}$.}
\label{fig:P}
\end{center}
\end{figure}

 In Fig \ref{fig:P} we show $P(k,a=1)$ and 
\begin{equation}
\left({\Delta P \over P}\right)(k,a=1)={[P(k,a=1)-P^{(0)}(k,a=1)] \over P^{(0)}(k,a=1)}=2\epsilon \mathcal{A}(k,a=1)\, ,
\end{equation}
which measures the relative deviation from the matter power spectrum in a $\Lambda$CDM model. 

At this point, let us stress that, knowing 
\begin{equation}
{\partial f \over \partial R}=1+\epsilon \xi
\end{equation}
and defining Eq. (\ref{Bxi}) today, i.e.
 \begin{equation}
 \label{Bxi0}
B_0=\frac{1}{\left(1-\Omega_{0 \, \rm m} + \Omega_{0 \, \Lambda}/2\right)} \left(- \epsilon \frac{\ud  \xi}{\ud a} \right)\bigg|_{a=1} \;,
 \end{equation}
we can compare a generic model in the literature characterized by $B_0$ and $f_{R0}=(\partial f/\partial R)_0$ with our approach (see the discussion in Section~\ref{FLRW}).
In particular we make a comparison of our results  with  the Hu-Sawicki model, see for example \cite{Hu:2007nk}.
In this case, using their definitions,  we get
\begin{equation}
R+\tilde{f}(R)=-2\Lambda-\frac{\tilde{f}_{R0}}{n}\frac{R_0^{n+1}}{R^n}\;,
\end{equation}
where $\tilde{f}(R)=f(R)-R$ and $\tilde{f}_{R0}=(\partial \tilde{f}/\partial R)_0$ in the notation of \cite{Hu:2007nk}.
Here we have used an approximate definition of their Lagrangian, i.e. when this model describes a background similar to $\Lambda$CDM.
Then $\epsilon \xi(a=1)=\tilde{f}_{R0}$ and find
\begin{equation}
\frac{\ud  \xi}{\ud a}\bigg|_{a=1}=(n+1)\frac{ \left(1-\Omega_{0 \, \rm m} + \Omega_{0 \, \Lambda}/2\right)}{\left(1+\Omega_{0 \, \rm m} - \Omega_{0 \, \Lambda}/2\right)}\xi|_{a=1}\;.
\end{equation}
Using the initial solutions as in Section \ref{FLRW} we see that we are considering a Hu-Sawicki model  with $n=0.82855$.
Finally, from Fig \ref{fig:P}  we can conclude that our technique work for $|\tilde{f}_{R0}|  < 10^{-8}$.
Obviously, in order to get a more precise result we have to add $\epsilon^2$-order contributions.

\section{Conclusions} \label{conclusions}

In this paper we have introduced a novel method which allows to describe the background evolution as well as cosmological perturbations in 
$f(R)$ modified gravity models, under the main assumption that the background evolution is close to $\Lambda$CDM. Here we restricted our analysis to linear perturbations, 
although the method is completely general and can be extended to any perturbative order.  

Let us conclude by adding some comments. As we stressed before, this approach is completely different with respect to previous ones. 
Indeed, through this analytic technique we can determine quantitatively the deviation in the behavior of cosmological perturbations between a given 
$f(R)$ model and $\Lambda$CDM reference model. Moreover, our treatment is general in that all the results depend only on the initial conditions that characterize the type of $f(R)$ model we are studying. Specifically all our expressions are completely determined by using the well-known results of the $\Lambda$CDM model (see Appendix \ref{A}) and the dynamical solution of $\xi$.

Our method allows to study structure formation in these models from the largest scales, of the order of the Hubble horizon, down to scales deeply inside the Hubble radius, without employing 
the so-called ``quasi-static" approximation \cite{Tsujikawa:2007gd} (see also the review \cite{DeFelice:2010aj} and refs. therein). 
This can be very useful as: 1) one can use this approach including all horizon-scale corrections to correctly interpret data on very large scales 
\cite{Yoo:2010jd, Yoo:2010ni, Bonvin:2011bg, Challinor:2011bk, Jeong:2011as}; 
2) at the non-linear level, it can be used to get  the expressions for the effect of primordial non-Gaussianity on the matter density perturbation in an $f(R)$ 
cosmology, fully accounting for the corrections arising on scales comparable with the Hubble radius~\cite{second-order} (e.g. in $\Lambda$CDM 
cosmology see \cite{Bartolo:2005xa, Bartolo:2010rw, Bartolo:2010ec, Bruni:2011ta, Baldauf:2011bh, Yoo:2011zc}). 

We believe that this technique can be extended to other classes of modified theories of gravity to providing information about the evolution of large-scale structures in the universe.

\vspace{0.5in}

\acknowledgments{DB would like to acknowledge the ICG at the University of Portsmouth for hospitality during the development of 
this project. DB research has been partly supported by ASI contract I/016/07/0 ``COFIS". 
Support was given by the Italian Space Agency through the ASI contracts Euclid-IC (I/031/10/0). We thank  V.~Acquaviva, D.~Bacon, B.~Hu, K.~Koyama, L.~Pogosian, A.~Raccanelli, F.~Schmidt, G.~Zhao for helpful discussions.}

\appendix
\section{Evolution of first-order perturbations of $\Lambda$CDM model in the synchronous gauge}\label{A}

In this section let us consider briefly the evolution, at the linear level, of relativistic perturbations of a $\Lambda$CDM model in the synchronous gauge.
Starting from the line-element defined in section \ref{Sec:PreRel},  in a  perfectly  homogeneous and  isotropic universe, the momentum constraint, 
the continuity equation and the energy constraint, respectively, give\footnote{For simplicity, in these appendices, we have substituted the double superscript $(1,0)$ with $(1)$.}
 \begin{eqnarray}
 && \psi^{(1)}+\frac{1}{6} \nabla^2 \chi^{(1)}= \psi_0^{(1)}+\frac{1}{6} \nabla^2 \chi_0^{(1)}= {\rm const.} \;, \\
 && \delta^{(1)}=\delta_0^{(1)} - \frac{1}{2} \nabla^2 \left(\chi^{(1)} - \chi_0^{(1)}\right) \;, \\
 && \nabla^2 \left[ 
\mathcal{H} \chi^{(1)}\,' + 4 \pi a^2 \rho^{(0)}_{\rm m}  \left( \chi^{(1)} -  \chi^{(1)}_{0} \right)  + 2 \psi^{(1)}_{0} + \frac{1}{3} \nabla^2 \chi^{(1)}_{0} 
\right] = 8 \pi a^2 \rho^{(0)}_{\rm m}  \delta^{(1)}_0 \;.
 \end{eqnarray}
The evolution equation becomes 
\begin{equation}\label{EE}
 \chi^{(1)}\,'' + 2 \mathcal{H}  \chi^{(1)}\,'+   \frac{1}{3}  \nabla^2 \chi^{(1)} = -2 \psi^{(1)} \;.
\end{equation}
An equation only for the scalar mode $\chi^{(1)}$  can be obtained by 
combining together the evolution equation and the energy constraint,
\begin{equation}\label{EC}
\nabla^2 \left[ \chi^{(1)}\,'' + \mathcal{H} \chi^{(1)}\,'  -4 \pi a^2 \rho^{(0)}_{\rm m}  \left( \chi^{(1)} - \chi^{(1)}_0 \right) \right] = - 
8 \pi a^2 \rho^{(0)}_{\rm m}  \delta^{(1)}_0 \;,
\end{equation}
and, consequently, we can get the equation for the linear density fluctuation  
\begin{equation}
{\delta^{(1)}}\,'' + \mathcal{H} \delta^{(1)}\,' - 4 \pi a^2 \rho^{(0)}_{\rm m} \delta^{(1)} = 0 \;.
\end{equation}
The equations above have been obtained in whole generality; 
one could have used 
instead the well-known residual gauge ambiguity of the synchronous 
coordinates (see, e.g., Refs.\cite{Matarrese:1995sb, Russ:1995eu, Matarrese:1997ay}) to simplify 
their form. 
For instance, one could fix $\chi^{(1)}_0$ so that  $\nabla^2 \chi^{(1)}_{0} = - 2 \delta^{(1)}_0$, and thus the 
$\chi^{(1)}$ evolution equation takes the same form as that for $\delta^{(1)}$, i.e. $\delta^{(1)}=-(1/2)\nabla^2 \chi^{(1)}$. 
Now, this gauge-condition replaced in Eqs.\ (\ref{EE}) and (\ref{EC}) yields
\begin{equation}
\label{***}
 \mathcal{H} \chi^{(1)}\,'   + 4 \pi a^2 \rho^{(0)}_{\rm m}   \chi^{(1)} + \frac{1}{3}  \nabla^2 \chi^{(1)} = -2 \psi^{(1)}
\end{equation}
Moreover, from the momentum constraint we get $\psi^{(1)}=\psi_0^{(1)} - (1/6) \nabla^2 (\chi^{(1)} - \chi_0^{(1)})$.
Finally with such a gauge fixing one obtains $\delta^{(1)} \propto \mathcal{D}_{\pm}$, where 
$\mathcal{D}_{\pm}$ represent the the growing ($+$) and decaying ($-$) solution of the equation
\begin{equation}
\mathcal{D}_{\pm}\,'' + \mathcal{H} \mathcal{D}_{\pm}\,' - 4 \pi a^2 \rho^{(0)}_{\rm m} \mathcal{D}_{\pm} = 0 \;.
\end{equation}
In what follows, we shall restrict ourselves to the growing mode. Then
\begin{eqnarray}
\label{SOL}
\chi^{(1)}=\mathcal{D}_{+}(\eta) \chi_0^{(1)} \;; \quad   \quad {\rm and} \quad  \quad  \psi^{(1)}=\psi_0^{(1)} - \frac{1}{6} \left( \mathcal{D}_{+}(\eta) -1 \right) \nabla^2  \chi_0^{(1)}\;,
\end{eqnarray} 
where we have defined $\mathcal{D}_{+}(\eta_0)=a_0=1$. Replacing (\ref{***}) in (\ref{SOL}) we find
\begin{equation}
\mathcal{H} ^2 \mathcal{D}_{+} \left[f(\Omega_{\rm m})+  \frac{3}{2} \Omega_{\rm m} \right]  \chi_0^{(1)} = - 2\left(\psi_0^{(1)}+\frac{1}{6} \nabla^2 \chi_0^{(1)} \right) = {\rm const.}
\end{equation}
where $ \Omega_{\rm m} =8 \pi  a^2 \rho^{(0)}_{\rm m} /(3 \mathcal{H}^2 )$ and\footnote{Specifically $f( \Omega_{\rm m} )=1 + g'/ (\mathcal{H}  g)$ where $g(z)=\mathcal{D}_{+}/a$ the growth suppression factor for a $\Lambda$CDM Universe (see Appendix \ref{B}).} 
$f( \Omega_{\rm m} )=\ud \ln \mathcal{D}_{+} / \ud \ln a$.
Then
\begin{eqnarray}
\psi^{(1)}=- \frac{1}{2}\mathcal{H}_0^2  \left[f(\Omega_{\rm 0 \, m})+  \frac{3}{2} \Omega_{\rm 0 \,  m} \right]   \chi_0^{(1)} - \frac{1}{6} \mathcal{D}_{+}(\eta) \nabla^2  \chi_0^{(1)}\;.
 \end{eqnarray}
It may be convenient to define the gravitational potential today $\Phi_0$ (see Appendix \ref{B}) through the relation $\nabla^2 \Phi_0=4 \pi  a_0^2 \rho^{(0)}_{\rm 0 \, m} \delta^{(1)}_0= -2\pi a_0^2 \rho^{(0)}_{\rm 0 \, m}  \nabla^2  \chi_0^{(1)} $. Then $ \chi_0^{(1)} = - 4\Phi_0/(3\mathcal{H}_0^2 \Omega_{\rm 0 \,  m})$. Finally we obtain
\begin{eqnarray}
  \psi^{(1)}= \left[1 + \frac{2}{3} \frac{f(\Omega_{\rm  0 \, m})}{\Omega_{\rm 0 \,  m}}\right]  \Phi_0 + \frac{2}{9} \frac{\nabla^2 \Phi_0}{\mathcal{H}_0^2 \Omega_{\rm 0 \, m}  } \mathcal{D}_{+} \;.
 \end{eqnarray}
 
\section{Evolution of first-order perturbations of $\Lambda$CDM model in the Poisson gauge}\label{B}
 
The goal of this Appendix is to briefly recall the results for the linear perturbations in 
the case of a non-vanishing cosmological $\Lambda$ term in the Poisson gauge~\cite{Bert, Mukhanov:1990me}.
 
Starting from the line-element defined in section \ref{poissongauge}, we note immediately that, at linear order, 
the traceless part of the ($i$-$j$)-components of Einstein's equations gives $\Phi^{(1)}=\Psi^{(1)} \equiv \Phi$. 
Its trace gives the evolution equation for the linear scalar potential $\Phi$
\begin{equation}
\label{ev}
\Phi''+3 {\mathcal H} \Phi'+ (2 \mathcal{H}' + \mathcal{H}^2) \Phi=0\, .
\end{equation}
Selecting only the growing mode solution one can write
\begin{equation}
\label{relphiphi_0}
\Phi({\bf x}, \eta)= g(\eta)\, \Phi_0({\bf x}) \, ,
\end{equation}
where $\Phi_0$ is the peculiar gravitational potential, linearly 
extrapolated to the present time $\eta_0$ and $g(\eta)=D_+(\eta)/a(\eta)$ is the so called 
growth-suppression factor.
The exact form of $g$ can be found in 
Refs.~\cite{Lahav:1991wc,Carroll:1991mt,Eisenstein:1997ij}. 
In the $\Lambda=0$ case, $g=1$. A very good approximation for $g$ as a 
function of redshift $z$ is given in 
Refs.~\cite{Lahav:1991wc,Carroll:1991mt} 
\begin{equation}
g \propto  \Omega_{\rm m} \left[ \Omega_{\rm m}^{4/7} - \Omega_\Lambda +
\left(1+  \Omega_{\rm m} /2\right)\left(1+ \Omega_\Lambda/70\right)\right]^{-1} \;, 
\end{equation}
with $ \Omega_{\rm m}=\Omega_{\rm 0 \,  m}(1+z)^3/E^2(z)$, 
$\Omega_\Lambda=\Omega_{0\Lambda}/E^2(z)$, 
$E(z) \equiv (1+z) {\mathcal H}(z)/{\mathcal H}_0 = [\Omega_{\rm 0 \,  m}(1+z)^3 + 
\Omega_{0 \, \Lambda}]^{1/2}$ and 
$\Omega_{\rm 0 \,  m}$, $\Omega_{0 \, \Lambda}=1-\Omega_{\rm 0 \,  m}$, the present-day
density parameters of non-relativistic matter and cosmological constant, 
respectively. According to our normalization, 
$g(z=0)=1$. 
The energy and momentum constraints provide the density and
velocity fluctuations in terms of $\Phi$ 
(see, for example, Ref.~\cite{Bartolo:2003gh} and~\cite{Mollerach:2003nq,Tomita:2005et} 
for the $\Lambda$ case)
\begin{eqnarray} 
\label{lindensvel}
\delta^{(1)}_{\rm p}& = & \frac{1}{4 \pi a^2 \rho_{\rm m}^{(0)}} 
\left[ \nabla^2 \Phi - 3{\cal H}\left(\Phi' + {\cal H} 
\Phi\right) \right], \\
\label{v1}
v^{(1)}_{\rm p}& = & - \frac{1}{4 \pi a^2 \rho_{\rm m}^{(0)}} 
\left(\Phi' + {\cal H} \Phi\right) \;.  
\end{eqnarray}

\bibliographystyle{JHEP}


\end{document}